\documentstyle[epsf]{article}

\newcommand{\ct}{\cite}
\newcommand{\lb}{\label}

\newcommand{\bc}{\begin{center}}
\newcommand{\ec}{\end{center}}
\newcommand{\bd}{\begin{displaymath}}
\newcommand{\ed}{\end{displaymath}}
\newcommand{\be}{\begin{equation}}
\newcommand{\ee}{\end{equation}}
\newcommand{\ba}{\begin{array}}
\newcommand{\ea}{\end{array}}
\newcommand{\bt}{\begin{tabular}}
\newcommand{\et}{\end{tabular}}

\newcommand{\bp}{\begin{picture}}
\newcommand{\ep}{\end{picture}}
\newcommand{\bfi}{\begin{figure}}
\newcommand{\efi}{\end{figure}}

\def\fun#1#2{\lower3.6pt\vbox{\baselineskip0pt\lineskip.9pt
\ialign{$\mathsurround=0pt#1\hfil##\hfil$\crcr#2\crcr\sim\crcr}}}

\begin{document}

\vspace{3cm}

\title{\Large\bf { Dirac Relation and Renormalization Group Equations
for Electric and Magnetic Fine Structure Constants }}
\author{{\bf L.V.Laperashvili}
\footnote{{\bf E-mail}:laper@heron.itep.ru, larisa@vxitep.itep.ru}\\
\it Institute of Theoretical and Experimental Physics,\\
\it B.Cheremushkinskaya 25, 117218 Moscow, Russia \\[0.2cm]
{\bf H.B.Nielsen}
\footnote{{\bf E-mail}:hbech@nbivms.nbi.dk, Holger.Bech.Nielsen@cern.ch, adress from 01.08.1999 to  01.08.2000: Theory Division CERN Ch 1211 Geneva 23 Switzerland Telephone +41 22 7678757}\\
\it Niels Bohr Institute,\\
\it DK-2100, Copenhagen, Denmark;\\
\it Theory Division, CERN \\
\it ch 1211 Geneva 23, Switzerland }

\date{}

\maketitle

\vspace{1cm}

{\large
 Dirac relation, renormalization, duality  \\

\vspace{3cm}


\begin{abstract}
The quantum field theory describing electric and magnetic charges
and revealing a dual symmetry was developed in the Zwanziger formalism.
The renormalization group (RG) equations
for both fine structure constants -- electric $\alpha $ and magnetic
$\tilde \alpha $ -- were obtained. It was shown that the Dirac relation
is valid for the renormalized $\alpha $ and $\tilde \alpha $ at the
arbitrary scale, but these RG equations can be considered perturbatively
only in the small region:
$0.25 \stackrel{<}{\sim} \alpha , \tilde \alpha \stackrel{<}{\sim} 1$
with $\tilde \alpha $ given by the Dirac relation: $\alpha {\tilde
\alpha} = \frac 14 $.
\end{abstract}

\newpage

\pagenumbering{arabic}
\vspace{1cm}
\large
\section{ Introduction}
\vspace{0.51cm}

The existence of the renormalization group (RG) in the quantum field
theory was discovered by E.C.G.Stueckelberg and A.Peterman \ct{1}.
RG techniques were successfully developed by Gell--Mann and Low \ct{2} in
their investigation of the effective charge behavior. They first
noticed that the derivative $\large d\log \alpha (p)/dt$ is only a
function of the effective fine structure constant:
\be
   \alpha(p) = \frac {e^2(p)}{4\pi}     \lb{1}
\ee
where $e(p)$ is the effective charge, $p$ is a 4--momentum and
\be
          t = \log \frac {p^2}{M^2}      \lb{2}
\ee
with $M$ as a momentum cut--off.

In gauge theories without monopoles the Gell--Mann--Low RG equation has
the following form:
\be
          \frac {d\log \alpha(p)}{dt} = \beta (\alpha(p))    \lb{3}
\ee
where the function $\beta(\alpha)$ depends on the Lagrangian describing
the theory.

At sufficiently small charge $(\alpha < 1)$ the $\beta$--function is
given by a series over $\alpha /4\pi$:
\be
\beta(\alpha) = \beta_2(\frac{\alpha}{4\pi}) +
\beta_4{(\frac{\alpha}{4\pi})}^2 + ...               \lb{4}
\ee
The first two terms of this series were calculated in QED a long time
ago [3,4]. The following result was obtained in the framework of the
perturbation theory (in the one-- and two--loop approximations):

a)
\be
      \beta_2 = \frac 43, \quad \quad \beta_4 = 4 \quad \quad -\quad
\mbox{ for fermion (electron) loops}
                                                     \lb{5}
\ee
and

b)
\be
     \beta_2 = \frac 13, \quad \quad \beta_4 = 1 \quad \quad -\quad
\mbox{ for scalar particle loops}.
                                                      \lb{6}
\ee
This result means that for both cases a) and b) the $\beta$--function
can be represented by the following series:
\be
       \beta(\alpha) = \beta_2(\frac{\alpha}{4\pi})(1 + 3\frac{\alpha}
                        {4\pi} + ...)
                                                \lb{7}
\ee
and we are able to use the one--loop approximation up to $\alpha \sim 1$
(with accuracy $\approx {25\%}$ for $\alpha \approx 1$).

In the present paper we consider the Abelian quantum field theory
when both electrically and magnetically charged particles with
charges $e$ and $g$, respectively, present in the theory which we
call below QEMD ("quantum electromagnetodynamics") following
the terminology used in Ref.\ct{5}. In their review \ct{5}, M.Blagojevic
and P.Senjanovic described the various formulations [6-15] of the Abelian
quantum field theories containing two coupling constants (electric and
magnetic charges) connected via the quantization condition. Several topics
were treated there: Dirac's and Schwinger's quantum mechanics of the
monopole, connection with non-Abelian monopoles, a supersymmetric
generalization of the theory and other aspects.

The aim of this paper is to investigate in QEMD the corresponding
RG equations for electric $(\alpha)$ and
magnetic $(\tilde{\alpha}) $ fine structure constants
in accordance with the Dirac relation for the minimal charges:
\be
              eg = 2\pi ,                  \lb{8}
\ee
or
\be
               \alpha \tilde{\alpha} = \frac 14    \lb{9}
\ee
where
\be
        \tilde \alpha = \frac {g^2}{4\pi}.              \lb{10a}
\ee
Below we consider QEMD in the Zwanziger formalism \ct{9}.

\section{The Zwanziger formalism for the Abelian gauge theory with electric
and magnetic charges}
\vspace{0.51cm}

A version of the local field theory for the Abelian gauge fields interacting
with electrically and magnetically charged particles is represented by the
Zwanziger formalism [9,10] (see also \ct{16}) which considers two potentials
$A_{\mu}(x)$ and $B_{\mu}(x)$ describing one physical photon with two
physical degrees of freedom.

In this theory the total field system of the gauge, electrically- ($\Psi$) and
magnetically-charged ($\Phi$) fields is described by the partition function
which has the following form in Euclidean space:
\be
 Z = \int [DA][DB][D\Phi ][D\bar{\Phi}][D\Psi ][D\bar{\Psi}]e^{-S} \lb{10}
\ee
where
\be
       S = S_{Zw}(A,B) + S_{gf} + S_e + S_m.
                                                       \lb{11}
\ee
The Zwanziger action $S_{Zw}(A,B)$ is given by:
$$
      S_{Zw}(A,B) = \int d^4x [\frac 12 {(n\cdot[\partial \wedge A])}^2 +
                  \frac 12 {(n\cdot[\partial \wedge B])}^2 +\\
$$
\be
     +\frac i2(n\cdot[\partial \wedge A])(n\cdot{[\partial \wedge B]}^*)
       - \frac i2(n\cdot[\partial \wedge B])(n\cdot{[\partial \wedge A]}^*)]
                                                              \lb{12}
\ee
where we have used the following designations:
$$
       {[A \wedge B]}_{\mu\nu} = A_{\mu}B_{\nu} - A_{\nu}B_{\mu},
\quad {(n\cdot[A \wedge B])}_{\mu} = n_{\nu}{(A \wedge B)}_{\nu\mu},\\
$$
\be
\quad G^*_{\mu\nu} = \frac 12\epsilon_{\mu\nu\lambda\rho}G_{\lambda\rho},
                                           \lb{13}
\ee
with $n^{\mu}$ representing the direction of the frozen Dirac string.

$S_{gf}$ in Eq.(\ref{11}) is the gauge--fixing action and the actions $S_e$
and $S_m$:
\be
           S_{e,m} = \int d^4x L_{e,m}(x)         \lb{14}
\ee
describe the matter fields with the electric and magnetic charges,
respectively.

Here we have a number of possibilities. The electrically- and 
magnetically-charged fields can be described by the following Lagrangian 
expressions
(given in Minkowski space):

\newpage

a)

\be
           L_e = \bar{\Psi}{\gamma}_{\mu}(iD_{\mu} - \mu_e)\Psi,
                                               \lb{15}
\ee
\be
           L_m = \bar{\Phi}{\gamma}_{\mu}(i{\tilde D}_{\mu} - \mu_m)\Phi
                                               \lb{16}
\ee
if they are fermions (electrons or monopoles, respectively).
In Eqs.(\ref{15}), (\ref{16})
\be
              D_{\mu} = \partial_{\mu} - ieA_{\mu}   \lb{17}
\ee
and
\be
       {\tilde D}_{\mu} = \partial_{\mu} - igB_{\mu}   \lb{18}
\ee
are the covariant derivatives.

b)

\be
       L_e = \frac 12 [{|D_{\mu}\Psi|}^2 - {\mu_e}^2{|\Psi|}^2],
                                                      \lb{19}
\ee
\be
       L_m = \frac 12 [{|{\tilde D}_{\mu}\Phi|}^2 - {\mu_m}^2{|\Phi|}^2]
                                                      \lb{20}
\ee
if the electrically- and magnetically-charged particles are the Klein--Gordon
complex scalars. But for the Higgs scalars with electric and magnetic charges
we have the following Lagrangians:

c)

\be
      L_e = \frac 12 {|D_{\mu}\Psi|}^2 - U(|\Psi|),
                                                   \lb{21}
\ee
\be
      L_m = \frac 12 {|\tilde D_{\mu}\Psi|}^2 - U(|\Phi|)
                                                   \lb{22}
\ee
where
\be
      U(|\Psi|) = \frac 12 \mu_e^2{|\Psi|}^2 + \frac{\lambda_e}4{|\Psi|}^4
                                                \lb{23}
\ee
and
\be
      U(|\Phi|) = \frac 12 \mu_m^2{|\Phi|}^2 + \frac{\lambda_m}4{|\Phi|}^4
                                                \lb{24}
\ee
are the Higgs potentials for the electrically- and magnetically-charged 
fields, respectively.
The complex scalar fields:
\be
      \Phi = \phi + i\chi_1 \quad\quad
   \mbox{and}\quad\quad \Psi = \psi + i\chi_2    \lb{24a}
\ee
contain Higgs ($\phi, \psi $) and Goldstone ($\chi_1, \chi_2 $) boson fields.

Below we shall consider the gauge-fixing action $S_{gf}$ chosen
in Ref.\ct{16}:
\be
     S_{gf} = \int d^4x [\frac{M^2_A}2{(n\cdot A)}^2
                         + \frac{M^2_B}2{(n\cdot B)}^2]
                                                     \lb{29}
\ee
which has no ghosts.

\section{Dual Symmetry and Charge Quantization Conditions}
\vspace{0.51cm}

In the last years gauge theories essentially operate with the
fundamental idea of duality (see, for example, reviews \ct{8} and
references there).

Duality is a symmetry appearing in free electromagnetism as invariance
of the free Maxwell equations:
\be
 \bf{\bigtriangledown}
\cdot \vec{\bf B} = 0, \quad\quad
 \bf{\bigtriangledown}\times \vec{\bf E} = - \partial_0\vec{\bf B},                 \lb{30}
\ee
\be
 \bf{\bigtriangledown}\cdot \vec{\bf E} = 0, \quad\quad
 \bf{\bigtriangledown}\times \vec{\bf B} = \partial_0\vec{\bf E},                 \lb{30a}
\ee
under the interchange of electric and magnetic fields:
\be
    \vec{\bf E} \to \vec{\bf B}, \quad\quad
        \vec{\bf B} \to  -\vec{\bf E}.                  \lb{31}
\ee
Letting
\be
      F = \partial\wedge A = - (\partial\wedge B)^{*},        \lb{32}
\ee
\be
      F^{*} = \partial\wedge B = (\partial\wedge A)^{*},        \lb{33}
\ee
it is easy to see that the following equations:
\be
    \partial_\lambda F_{\lambda\mu} = 0,                   \lb{34}
\ee
which together with the Bianchi identity:
\be
    \partial_\lambda F_{\lambda\mu}^{*} = 0                \lb{35}
\ee
are equivalent to Eqs.(\ref{30}), show invariance under the Hodge star
operation on the field tensor:
\be
     F_{\mu\nu}^{*} = \frac 12 \epsilon_{\mu\nu\rho\sigma} F_{\rho\sigma}
                                                  \lb{36}
\ee
(here $F^{**} = - F$).

This Hodge star duality applied to the free Zwanziger Lagrangian (\ref{12})
leads to its invariance under the following duality transformations:
\be
   F \leftrightarrow F^{*},\quad\quad
(\partial\wedge A) \leftrightarrow (\partial \wedge B),
\quad\quad (\partial\wedge A)^{*} \leftrightarrow - (\partial
                \wedge B)^{*}.
                                                \lb{37}
\ee
Introducing the interacting Maxwell equations:
\be
    \partial_\lambda F_{\lambda\mu} = j_{\mu}^e,                   \lb{38}
\ee
\be
    \partial_\lambda F_{\lambda\mu}^{*} = j_{\mu}^m,                \lb{39}
\ee
with the local conservation laws for the electric and magnetic charges:
\be
         \partial_{\mu}j_{\mu}^{e,m} = 0,            \lb{40}
\ee
we immediately see the invariance of these equations under the exchange
of the electric and magnetic fields (Hodge star duality) provided that
at the same time electric and magnetic charges and currents
(and masses if they are different) are also interchanged:
\be
     e \leftrightarrow g, \quad\quad j_{\mu}^e \leftrightarrow j_{\mu}^m
                  \lb{41}
\ee
together with $\mu_e \leftrightarrow \mu_m$ in Eqs.(\ref{15}), (\ref{16}),
(\ref{19}), (\ref{20}), (\ref{23}), (\ref{24}),
and $\lambda_e \leftrightarrow \lambda_m$ in Eqs.(\ref{23}), (\ref{24}).

The corresponding quantum field theory with electric $e_i$ and magnetic
$g_i$ charges is selfconsistent if both charges are quantized according
to the famous Dirac relation \ct{6}:
\be
               e_ig_j = 2\pi n_{ij}      \lb{42}
\ee
when $n_{ij}$ is an integer. For $n_{ij} = 1$ we have the Dirac
quantization condition (\ref{8}) in terms of the elementary electric and
magnetic charges. But J.Schwinger \ct{8} showed another possibility
for the charge quantization condition when instead of Eq.(\ref{8})
we have:
\be
                   eg = 4\pi.        \label{42a}
\ee
The results (\ref{8}) and (\ref{42a}) depend on the choice of the string
singularity  line. We prefer to consider the Dirac semi-infinite
string and the Dirac relation (\ref{8}) as a charge quantization condition.

If the fundamental electric charge $e$ is so small that it corresponds to
the perturbative electric theory, then magnetic charges are large and
correspond to the strongly interacting magnetic theory, and vice versa.
But below we consider some small region of $e,g$ values
(we hope that it exists) which allows us to employ the
perturbation theory in both, electric and magnetic, sectors.

When non-trivial dyons -- particles with both electric and magnetic charges
simaltaneously -- are present, then the analogue of the Dirac relation
becomes a little more complicated and it then reads:
\be
          e_ig_j - e_jg_i = 2\pi n_{ij}     \lb{43}
\ee
which is duality invariant (see for example the reviews \ct{5}, \ct{17} and
the references there).

The relation (\ref{43}) has the name of the Dirac--Schwinger--Zwanziger
[6,8,9] quantization condition. But in this paper the theory of dyons
is not exploited.

\section{Propagators}
\vspace{0.51cm}

At the same time as the partition function (\ref{10}) let us consider
the generating functional with external sources
$J_{\mu}^{(A)}, J_{\mu}^{(B)}, \eta$ and $\omega$:
$$
        Z[J^{(A)}, J^{(B)}, \eta, \omega ] =
$$
\be
= \int [DA][DB][D\Phi ][D\bar{\Phi}][D\Psi ][D\bar{\Psi}] e^{ - S
       + (J^{(A)},A) + (J^{(B)},B) + (\bar{\eta},\Phi ) + (\bar{\Phi},\eta )
       + (\bar{\omega},\Psi ) + (\bar{\Psi},\omega )}
                                                          \lb{44}
\ee
where
\be
      (J,A) = \int d^4x J_{\mu}(x)A_{\mu}(x),\quad \mbox{and}\quad
      (\bar{\eta},\Phi ) = \int d^4x \bar{\eta}(x)\Phi (x),\quad \mbox{etc.}
                                                            \lb{45}
\ee
Using this generating functional it is not difficult to
calculate the propagators of the fields considered in our model.

Three "bare" propagators of the gauge fields $A_{\mu}$ and $B_{\mu}$:
$$
 Q_{\mu\nu}^{0(A)} = < A_{\mu}A_{\nu} > =
          \frac{\delta^2Z[J^{(A)},J^{(B)},\eta ,\omega ]}
          {\delta J_{\mu}^{(A)}\delta J_{\nu}^{(A)}},
$$
$$
      Q_{\mu\nu}^{0(B)} = < B_{\mu}B_{\nu} > =
                \frac{\delta^2Z[J^{(A)},J^{(B)},\eta ,\omega ]}
                        {\delta J_{\mu}^{(B)}\delta J_{\nu}^{(B)}},
$$
\be
Q_{\mu\nu}^{0(AB)} = < A_{\mu}B_{\nu} > =
               \frac{\delta^2Z[J^{(A)},J^{(B)},\eta ,\omega ]}
                      {\delta J_{\mu}^{(A)}\delta J_{\nu}^{(B)}}
                                                             \lb{46}
\ee
are presented in Fig.1 (see Fig.1(a)) together with the propagators
$Q_{\mu\nu}^{(A)}$, $Q_{\mu\nu}^{(B)}$ and
$Q_{\mu\nu}^{(AB)}$ determined by Fig.1(b).

Propagators (\ref{46}) were calculated by authors of
Ref.\ct{16} in the momentum space:
\be
      Q_{\mu\nu}^{0(A,B)}(q) = \frac 1{q^2}(\delta_{\mu\nu} +
     \frac{q^2 + {M_{A,B}}^2}{M^2_{A,B}}\frac{q_{\mu}q_{\nu}}{{(n\cdot q)}^2}
      - \frac 1{(n\cdot q)}(q_{\mu}n_{\nu} + q_{\nu}n_{\mu})),
                                                   \lb{47}
\ee
\be
     Q_{\mu\nu}^{0(AB)} = \frac {i}{q^2}\epsilon_{\mu\nu\rho\sigma}
\frac{q_{\rho}n_{\sigma}}{(n\cdot q)}.
                                                  \lb{48}
\ee

The dot on the diagrams of Fig.1(a,b) corresponds to the following operator:
\be
\Lambda_{\mu\nu} = iq^2\epsilon_{\mu\nu\rho\sigma}\frac{q_{\rho}
                         n_{\sigma}}{(n\cdot q)}.
                                                   \lb{49}
\ee
Considering QED with Lagrangian (\ref{15}) for electrons (monopoles are
absent in this case) it is easy to see that the "bare" $D_{\mu\nu}^0(q^2)$
and the "dressed" (renormalized) $D_{ren,\mu\nu}(q^2)$ photon propagators
obey the following relation presented in Fig.2:
\be
     D_{ren,\mu\nu}(q) = D_{\mu\nu}^0(q) +
          D_{\mu\kappa}^0(q) \Pi_{\kappa\lambda}(q)D_{ren,\lambda\nu}(q).
                                                     \lb{50}
\ee
Here the contribution of the electron loop is described by the
operator $\Pi_{\kappa\lambda}(q)$ given by the following expression:
\be
  \Pi_{\kappa\lambda}(q) = e^2\int \frac{d^4k}{{(2\pi )}^4}
 Tr[\gamma_{\kappa}G(k)\Gamma_{\lambda}(k, k - q)G(k - q)]
                                              \lb{51}
\ee
where $\gamma_{\kappa}$ is the Dirac matrix, $\Gamma_{\lambda}$
is the renormalized vertex and $G(k)$ is the "dressed" propagator of
the electron. Taking into account the transversality of the photon
self-energy tensor we have:
\be
     \Pi_{\kappa\lambda}(q)  = (q^2\delta_{\kappa\lambda}
               - q_{\kappa}q_{\lambda})\Pi(q^2).
                                      \lb{52}
\ee
The "dressed" propagators $Q_{ren,\mu\nu}^{(A)}$, $Q_{ren,\mu\nu}^{(B)}$ and
$Q_{ren,\mu\nu}^{(AB)}$ containing the contributions of the electric
(thin lines) and magnetic (thick lines) charged particle loops are
presented in Fig.3.

The Lagrangians (\ref{15}) and (\ref{16}) contain the interaction terms
$j_{\mu}^eA_{\mu}$ and $j_{\mu}^mB_{\mu}$ where $j_{\mu}^e$ and $j_{\mu}^m$
are the electric and magnetic currents:
\be
         j_{\mu}^e = e\bar{\Psi}\gamma_{\mu}\Psi       \lb{53}
\ee
and
\be
         j_{\mu}^m = g\bar{\Phi}\gamma_{\mu}\Phi.       \lb{54}
\ee
The interactions in the Lagrangians (\ref{19})-(\ref{22}) are given
by $j_{\mu}^eA_{\mu}$ and $j_{\mu}^mB_{\mu}$ as well as by "seagull"
terms $e^2A_{\mu}A_{\mu}\Psi^{+}\Psi$ and $g^2B_{\mu}B_{\mu}\Phi^{+}\Phi$,
respectively, but now we have:
\be
     j_{\mu}^e = e(\Psi^{+}\partial_{\mu} \Psi -
                              \Psi \partial_{\mu}\Psi^{+})  \lb{55}
\ee
and
\be
     j_{\mu}^m = g(\Phi^{+}\partial_{\mu} \Phi -
                               \Phi \partial_{\mu}\Phi^{+}).  \lb{56}
\ee

The simplest electron-electron, monopole-monopole and electron-monopole
interactions corresponding to the Lagrangians $L_{e,m}$ given
by Eqs.(\ref{15}) and (\ref{16}) are shown in Fig.4.

Now we are ready to obtain the renormalization group equations
for the effective fine structure constants when both, electric and
magnetic charges are present in our field system.

\section{Renormalization Group Equations for the Electric and Magnetic
Fine Structure Constants}
\vspace{0.51cm}

The Gell--Mann--Low RG equation (\ref{3}) can be obtained by the calculation
of the "dressed" propagators. In the case of QED the relation shown in
Fig.2 gives us:
\be
             D(q) = Z_3^{-1}D^{0}(q)         \lb{57}
\ee
where the renormalization constant $Z_3$ is related, in its turn,
with $\Pi(q^2)$ determined by Eqs.(\ref{51}),(\ref{52}):
\be
        Z_3 = 1 - \Pi(\mu^2).
                                        \lb{58}
\ee
Here $\mu $ is the energy scale: $q^2 = \mu^2$.

The Gell--Mann--Low $\beta$-function is given by the following expression:
\be
  \beta(\alpha(p))
         = - \frac{\partial \log{Z_3}}{\partial \log \mu^2},
                                                     \lb{59a}
\ee
or
\be
  \beta(\alpha(p))
     = - \frac{\partial \log{(1 - \Pi(\mu^2))}}{\partial \log \mu^2}.
                                             \lb{59b}
\ee
In the one--loop approximation of perturbation theory
(see for example \ct{18}) we can write:
\be
  \beta(\alpha(p))
              \approx \frac{\partial \Pi(\mu^2)}{\partial \log \mu^2}
              \approx - \frac{\partial \Pi(M^2)}{\partial \log M^2}
                                                            \lb{59c}
\ee
where M is the momentum cut--off.

Let us consider now the renormalization group equations for $\alpha $
and $\tilde {\alpha}$ when both (electric and magnetic) charges
are present in our theory.

Fig.3 shows the contributions of the electric (thin lines) and magnetic
(thick lines) charged-particle loops to the "dressed" propagators
$Q_{ren,\mu\nu}^{(A)}$ and $Q_{ren,\mu\nu}^{(B)}$. We consider also
the "dressed" propagators in loops assuming the theory of the case a) with
Lagrangians $L_{e,m}$ given by Eqs.(\ref{15}) and (\ref{16}).

Introducing the renormalization constants $Z_3$ and $\tilde Z_3$ by
the following relations:
\be
     Q_{ren,\mu\nu}^{(A)} = Z_3^{-1} Q_{\mu\nu}^{(A)},  \lb{60}
\ee
\be
      Q_{ren,\mu\nu}^{(B)} = \tilde Z_3^{-1} Q_{\mu\nu}^{(B)},  \lb{61}
\ee
\be
      Q_{ren,\mu\nu}^{(AB)} =
{(Z_3 {\tilde Z}_3)}^{-1/2} Q_{\mu\nu}^{(AB)},  \lb{62}
\ee
it is not difficult to calculate $Z_3$ and  $\tilde Z_3$ according to the
diagrams shown in Fig.3. These diagrams demonstrate the following
relations:
\be
      Q_{ren,\mu\nu}^{(A,B)} = Q_{\mu\nu}^{(A,B)} +
 Q_{\mu\kappa}^{(A,B)}\Pi_{\kappa\lambda}^{(e,m)}Q_{ren,\lambda\nu}^{(A,B)} +
 Q_{\mu\kappa}^{(AB)}\Pi_{\kappa\lambda}^{(m,e)}Q_{ren,\lambda\nu}^{(AB)}
                                     \lb{62a}
\ee
which mean:
\be
        Z_3^{-1}Q_{\mu\nu}^{(A)} = Q_{\mu\nu}^{(A)} +
Q_{\mu\kappa}^{(A)}\Pi_{\kappa\lambda}^{(e)} Z_3^{-1} Q_{\lambda\nu}^{(A)} +
Q_{\mu\kappa}^{(AB)}\Pi_{\kappa\lambda}^{(m)} {(Z_3{\tilde Z}_3)}^{-1/2}
Q_{\lambda\nu}^{(AB)},
                              \lb{62b}
\ee
\be
     {\tilde Z}_3^{-1}Q_{\mu\nu}^{(B)} = Q_{\mu\nu}^{(B)} +
Q_{\mu\kappa}^{(B)}\Pi_{\kappa\lambda}^{(m)} {\tilde Z}_3^{-1}
Q_{\lambda\nu}^{(B)} +
Q_{\mu\kappa}^{(AB)}\Pi_{\kappa\lambda}^{(e)}{(Z_3{\tilde Z}_3)}^{-1/2}
Q_{\lambda\nu}^{(AB)},
                              \lb{62c}
\ee

Using the expressions
(\ref{47}) and (\ref{48}) obtained in Ref.\ct{16} for propagators
$Q^{0(A,B,AB)}_{\mu\nu}$ and the limit $M_{A,B}\to \infty$
in Eq.(\ref{47}), it is possible to show that the following relations
are valid:
\be
    Z_3^{-1}Q_{\mu\nu}^{(A)} = Q_{\mu\nu}^{(A)} +
{\Pi}^{(e)} Z_3^{-1} Q_{\mu\nu}^{(A)} + {\Pi}^{(m)} {(Z_3\tilde Z_3)}^{-1/2}
Q_{\mu\nu}^{(A)},     \lb{63a}
\ee
\be
     \tilde Z_3^{-1}Q_{\mu\nu}^{(B)} = Q_{\mu\nu}^{(B)} +
{\Pi}^{(m)} \tilde Z_3^{-1} Q_{\mu\nu}^{(B)} + {\Pi}^{(e)}{(Z_3
\tilde Z_3)}^{-1/2}Q_{\mu\nu}^{(B)},     \lb{63b}
\ee
or
\be
     Z_3 = \frac{1 - \Pi^{(e)}(\mu^2)}{1
               - \Pi^{(m)}(\mu^2){(Z_3\tilde Z_3)}^{-1/2}},   \lb{63}
\ee
\be
  \tilde Z_3 = \frac{1 - \Pi^{(m)}(\mu^2)}{1
               - \Pi^{(e)}(\mu^2){(Z_3\tilde Z_3)}^{-1/2}}.   \lb{64}
\ee
Here
$\Pi^{(e)}(\mu^2)$ and $\Pi^{(m)}(\mu^2)$
are given at $q^2 = \mu^2$ by the photon self--energy tensors
$\Pi_{\mu\nu}^{(e)}$ and $\Pi_{\mu\nu}^{(m)}$ corresponding to the electron
and monopole loops, respectively (see bubbles in Fig.3):
\be
     \Pi_{\mu\nu}^{(e,m)}(q)  = (q^2\delta_{\mu\nu}
               - q_{\mu}q_{\nu})\Pi^{(e,m)}(q^2)
                                      \lb{65}
\ee
where $\Pi^{(e)}$ (or $\Pi^{(m)})$ is described by Eqs.(\ref{51}),(\ref{52})
with $e, \mu_e$  (or $g, \mu_m)$ as a charge and a mass of
the electron (or monopole).

J.Schwinger was the first (see Refs.\ct{8}) who investigated
the renormalization problem of the electric and magnetic charges in QEMD.

Considering the "bare" charges $e_0$ and $g_0$ and
the renormalized effective charges $e$ and $g$, we must distinguish
between two opposite cases. It was shown in Refs.[8,11,12]:
\be
              e/g = e_0/g_0,   \lb{66a}
\ee
while other authors [13,14] obtained the following result:
\be
                eg = e_0g_0.    \lb{66b}
\ee
S.Coleman \ct{15} analysed the case when the extended t'Hooft--Polyakov
monopole is shrunk to zero size. The effective theory describing the
interaction between such objects in QEMD tells us something about
the renormalization effects. The consistency condition gave the
following result:
\be
             eg = {(Z_3{\tilde Z}_3)}^{1/2}e_0g_0 = 2\pi,  \lb{66}
\ee
or
\be
              Z_3{\tilde Z}_3 = 1.          \lb{67}
\ee
This means that the Dirac relation (\ref{8}) is valid not only for
the "bare" charges $e_0$ and $g_0$, but also for the renormalized
effective charges $e$ and $g$.

We have actually already rederived this result in the Zwanziger
formalism, since we can obtain it by using (\ref{63}) and (\ref{64}).
Let us in fact multiply (\ref{63}) on both sides with
$1/{\sqrt{ Z_3 }} - \Pi^{(m)}(\mu^2)/{\sqrt {{\tilde Z}_3}}$
and (\ref{64}) by
$1/{\sqrt{{\tilde Z}_3 }} - \Pi^{(m)}(\mu^2)/{\sqrt{Z_3}}$
and then add the resulting equations. The last ones become by the
cancellation of $\Pi$ terms giving
\be
  \sqrt{Z_3} + \sqrt{{\tilde Z}_3} =
  1/\sqrt{Z_3} + 1/\sqrt{{\tilde Z}_3},      \lb{67a}
\ee
from which it is easily seen that Eq.(\ref{67}) follows.

Using this result we obtain the following important relations:
\be
   Z_3 = {{\tilde Z}_3}^{-1} = \frac{1 - \Pi^{(e)}(\mu^2)}{1
                        - \Pi^{(m)}(\mu^2)}.
                                           \lb{68}
\ee
RG equations for the fine structure constants $\alpha $ and
$\tilde \alpha$ immediately follow from Eq.(\ref{68}):
\be
\frac{\mbox{d}(\log \alpha(p))}{\mbox{d}t} =
   - \frac{\partial \log{Z_3}}{\partial \log \mu^2},  \lb{69a}
\ee
or
$$
             \frac{\mbox{d}(\log \alpha(p))}{\mbox{d}t} =
    - \frac{\partial \log (1 - \Pi^{(e)} (\mu^2))}{\partial \log \mu^2}
     + \frac{\partial \log (1 - \Pi^{(m)} (\mu^2))}{\partial \log \mu^2}\\
$$
\be
     = \beta^{(e)}(\alpha) - \beta^{(m)}(\tilde{\alpha})
                                            \lb{69}
\ee
and
\be
\frac{\mbox{d}(\log \tilde \alpha(p))}{\mbox{d}t}
  = - \frac{\partial \log{\tilde Z}_3}{\partial \log \mu^2}
       = \frac{\partial \log{Z_3}}{\partial \log \mu^2}
       = \beta^{(m)}(\tilde {\alpha}) - \beta^{(e)}(\alpha ).
                                     \lb{70}
\ee
Here the analytical expressions for $\beta$--functions
are given by the same Eq.(\ref{59b}):
\be
\beta^{(e,m)}(\alpha \; \mbox {or}\; \tilde{\alpha})
     = - \frac{\partial \log{(1 - \Pi^{(e,m)}(\mu^2))}}{\partial \log \mu^2},
                                 \lb{70a}
\ee
but now these $\beta$--functions contain electron ($e,\mu_e$)
and monopole ($g,\mu_m$) parameters, respectively.

The obtained RG equations (\ref{69}) and (\ref{70}) obey the
following equality:
\be
 \frac{\mbox{d}(\log \alpha(p))}{\mbox{d}t} =
          - \frac{\mbox{d}(\log \tilde{\alpha}(p))}{\mbox{d}t}
                                          \lb{71}
\ee
which corresponds to the Dirac relation:
\be
  \alpha (t)\tilde{\alpha}(t) = \frac 14 \quad  \quad (\mbox{for all t})
                               \lb{72}
\ee
valid for the renormalized electric and magnetic fine structure constants
at the arbitrary scales.

\section{The beta--functions}

The functions $\beta^{(e,m)}$ are given perturbatively
by the expressions similar to Eq.(\ref{4}):
\be
\beta^{(e)}(\alpha) = \beta^{(e)}_2(\frac{\alpha}{4\pi}) +
\beta^{(e)}_4{(\frac{\alpha}{4\pi})}^2 + ...               \lb{4E}
\ee
and
\be
\beta^{(m)}(\tilde \alpha) = \beta^{(m)}_2(\frac{\tilde \alpha}{4\pi}) +
\beta^{(m)}_4{(\frac{\tilde \alpha}{4\pi})}^2 + ...               \lb{4M}
\ee
The perturbative expansions (\ref{4E}) and (\ref{4M}) coincide
with the series (\ref{4}) calculated in QED, at least on the level
of the two--loop approximation.
The monopole(electric) loops inside the electric(monopole) loops
appear only on the level of the three--loop approximation.
Of course, these $\beta$--functions are different if we
consider magnetic scalar particles instead of electric fermions, or
vice versa. The corresponding coefficients
$\beta^{(e)}_2,\;\beta^{(e)}_4$ of the series (\ref{4E})
or $\beta^{(m)}_2,\;\beta^{(m)}_4$ of the series
(\ref{4M}) are given by Eqs.(\ref{5}) or (\ref{6}) depending on the type of
the charged particles.

If both matter fields are electrically- and magnetically-charged fermions
or both are scalars then
we have the same expressions for the functions $\beta^{(e,m)}$
and we can write the following equations for the cases a) and b):
\be
             \frac{\mbox{d}(\log \alpha(p))}{\mbox{d}t} =
          - \frac{\mbox{d}(\log \tilde{\alpha}(p))}{\mbox{d}t}
            = \beta_2\frac{\alpha - \tilde \alpha}{4\pi}
               (1 + 3\frac{\alpha + \tilde \alpha}{4\pi} + ...).
                                      \lb{5B}
\ee
The last equations show that the one--loop approximation works with accuracy
$\stackrel{<}{\sim }30\% $ if both $\alpha$ and $\tilde \alpha$ obey
the following requirement:
\be
        0.25\stackrel{<}{\sim }
           \alpha, \tilde{\alpha} \stackrel{<}{\sim }1.    \lb{6B}
\ee
But strictly speaking, we don't know the exact behaviour of the whole
asymptotic series (\ref{5B}).

In Refs.\ct{19} and \ct{20} the behaviour of the effective fine structure
constants was investigated in the vicinity of the phase transition point
in compact (lattice) QED by the Monte--Carlo simulation method.
The following result was obtained [19,20]:
\be
     \alpha_{crit}^{lat}\approx 0.20\quad\quad {\mbox {and}}\quad\quad
    {\tilde \alpha}_{crit}^{lat}\approx 1.25.
                                                 \lb{7B}
\ee
These values almost coincide with the borders of the perturbation theory
requirement (\ref{6B}).
In consequence, assuming that the phase transition couplings (\ref{7B})
may be described by the one--loop approximation with accuracy not
worse than $(30-50)\%$, we have tried to calculate
them in the Higgsed monopole model (see Ref.\ct{21}).
The aim of the last paper was to confirm, in general, the idea of the
approximate "universality" (regularization independence) of the phase
transition couplings. The result obtained in \ct{21}:
\be
     \alpha_{crit}^{lat}\approx 0.18_5\quad\quad {\mbox {and}}\quad\quad
    {\tilde \alpha}_{crit}^{lat}\approx 1.35
                                                 \lb{8B}
\ee
is in accordance with the lattice result (\ref{7B}).
It seems that the idea of the approximate "universality" for the first-order 
phase transitions is really confirmed.

\section{Conclusions}
\vspace{0.51cm}

The aim of this paper was to obtain the renormalization group equations
for the electric and magnetic renormalized fine structure constants
using the Zwanziger formalism for QEMD.
The result (see Eqs.(\ref{69}) and (\ref{70})):
\be
       \frac{\mbox{d}(\log \alpha(p))}{\mbox{d}t} =
       -  \frac{\mbox{d}(\log \tilde \alpha(p))}{\mbox{d}t}
       = \beta^{(e)}(\alpha) - \beta^{(m)}(\tilde{\alpha})
                                           \lb{73a}
\ee
confirms the Dirac relation $\alpha (t)\tilde \alpha (t) =1/4$
existing at the arbitrary scale.

According to the philosophy given in the Introduction it is possible
to consider the perturbation theory for $\beta^{(e)}(\alpha )$ and
$\beta^{(m)}(\tilde{\alpha})$ simultaneously if both $\alpha $ and
$\tilde{\alpha}$ are sufficiently small. Then the functions $\beta^{(e,m)}$
are given perturbatively by the usual series (\ref{4}) or (\ref{7}).
The calculations in QED (see Section 1) have shown that the perturbation
theory works up to $\alpha \stackrel{<}{\sim } 1$. Due to the
Dirac relation (\ref{9}), such a requirement leads to the following
condition: $\tilde{\alpha}\stackrel{>}{\sim}0.25$.
In consequence, we have QEMD RG equations with beta-functions
$\beta^{(e,m)}$ considered perturbatively if both $\alpha $
and $\tilde {\alpha}$ obey the following requirement:
\be
        0.25\stackrel{<}{\sim }
          \alpha, \tilde{\alpha} \stackrel{<}{\sim }1.    \lb{73}
\ee
We have just this case in the vicinity of the phase transition point
for the compact (lattice) QED: $\alpha_{crit}^{lat}\approx{0.2}$
and ${\tilde{\alpha}}_{crit}^{lat}\approx{1.25}$ (see Refs.\ct{19}
and \ct{20}).

\vspace{0.5cm}

ACKNOWLEDGEMENTS: We would like to express special thanks to K.Milton
for his very fruitful criticism and advice, also to
D.L.Bennett and Ivan Shushpanov for useful discussions, and to
P.A.Kovalenko and D.A.Ryzhikh for their help in construction of
figures.

One of the authors (L.V.L.) is indebted to the Niels Bohr Institute
for its hospitality and financial support.

Financial support from grants INTAS-93-3316-ext and INTAS-RFBR-96-0567
is gratefully acknowledged.


\newpage
{\bf {\Large Captions for figures}:}

\vspace{1cm}

{\bf Fig.1}

(a) "Bare" propagators of gauge fields $A_{\mu}$
(thin wavy line) and $B_{\mu}$ (thin dashed line) describing a photon
in two (non--dual and dual) states existing in the Abelian gauge
theory with both electric and magnetic charges.\\

(b) Propagators of gauge fields $A_{\mu}$ (thick wavy line)
and $B_{\mu}$ (thick dashed line) describing the contribution
of dual transformations of the photon.

\vspace{1cm}

{\bf Fig.2}

"Dressed" propagator (double line) of the photon in QED.

\vspace{1cm}

{\bf Fig.3}

"Dressed" propagators of gauge fields $A_{\mu}$ (double wavy line)
and $B_{\mu}$ (double dashed line) containing the contributions
of the electrically- (thin lines) and magnetically-charged (thick lines) 
particle loops.

\vspace{1cm}

{\bf Fig.4}

The interaction of the electric (thin line) and magnetic (thick line)
currents.

\end{document}